\documentclass{ws-ijmpa}

\renewcommand{\bar}[1]{\overline{#1}}

\renewcommand{\d}{{\mathrm d}}

\begin{document}

\markboth{B.-Q.~Ma} {Nutev Anomaly Versus Strange-Antistrange
Asymmetry}

\catchline{}{}{}{}{}
%
%%%%%%%%%%%%%%%%%%%%%%%%%%%%%%%%%%%%%%%%%%%%%%%%%%%%%%%%%%%%%%%%%%%%

\title{NUTEV ANOMALY VERSUS STRANGE-ANTISTRANGE ASYMMETRY\footnote{Talk presented at International Conference on QCD and
Hadronic Physics, Beijing, China, June 16-20, 2005.}}

\author{\footnotesize BO-QIANG MA}

\address{School of Physics, Peking University, Beijing 100871, China}

\maketitle

%\pub{Received (Day Month Year)}{Revised (Day Month Year)}

\begin{abstract}
We report the correction from the asymmetric strange-antistrange
sea of the nucleon by using both the light-cone baryon-meson
fluctuation model and the chiral quark model, and show that a
significant part of the NuTeV anomaly can be explained by the
strange-antistrange asymmetry. We also show that the calculated
$s$/$\bar{s}$ asymmetry are compatible with the NuTeV data by
including some additional symmetric $s$/$\bar{s}$ quark
contribution.

\end{abstract}

%\vspace{0.32cm}
\vspace{0.28cm}

The NuTeV Collaboration\cite{zell02} at Fermilab measured the
value of the Weinberg angle (weak mixing angle)
$\sin^{2}\theta_{w}$ in deep inelastic scattering~(DIS) on nuclear
target with both neutrino and antineutrino beams. Having
considered and examined various source of systematic errors, the
NuTeV Collaboration reported the value:
$\sin^{2}\theta_{w}=0.2277\pm0.0013~(\mbox{stat})\pm0.0009~(\mbox{syst}),$
which is three standard deviations from the value
$\sin^{2}\theta_{w}=0.2227\pm0.0004$ measured in other electroweak
processes. As $\theta_{w}$ is one of the important quantities in
the standard model, this observation by NuTeV has received
attention by the physics society. This deviation, or NuTeV anomaly
as people called, could be an indication for new physics beyond
standard model, if it cannot be understood by a reasonable effect
within the standard model.

The NuTeV Collaboration measured the value of $\sin^{2}\theta_{w}$
by using the ratio of neutrino neutral-current and charged-current
cross sections on iron\cite{zell02}. This procedure is closely
related to the Paschos-Wolfenstein~(PW) relation\cite{pash73}:
\begin{equation}
R^{-}=\frac{\sigma^{{\nu}N}_{NC}-\sigma^{\overline{\nu}N}_{NC}}{\sigma^{{\nu}N}_{CC}
-\sigma^{\overline{\nu}N}_{CC}}=\frac{1}{2}-\sin^{2}\theta_{w},
\label{ratio}
\end{equation}
which is based on the assumptions of charge symmetry, isoscalar
target, and strange-antistrange symmetry of the nucleon sea.
%($s(x)\not=\bar{s}(x)$).
It is necessary to pay particular attention to the
strange-antistrange asymmetry\cite{bm97}, i.e.,
$s(x)\not=\bar{s}(x)$, which brings the correction to the PW
relation\cite{dm04}
\begin{equation}
R^{-}_{N}=\frac{\sigma^{\nu
N}_{NC}-\sigma^{\bar{\nu}N}_{NC}}{\sigma^{\nu
N}_{CC}-\sigma^{\bar{\nu}N}_{CC}} = R^{-}-\delta
R^{-}_{s},\label{correction}
\end{equation}
where $\delta R^{-}_{s}$ is the correction term
\begin{equation}
\delta
R^{-}_{s}=(1-\frac{7}{3}\sin^{2}\theta_{w})\frac{S^{-}}{Q_v+3
S^{-}},\label{rs}
\end{equation}
where $S^{-}\equiv\int^{1}_{0} x[s(x)-\bar{s}(x)]\textmd{d}x$ and
$Q_v \equiv\int^{1}_{0} x[u_{v}(x)+d_{v}(x)]\textmd{d} x$. I will
show in this talk that the effect due to the strange-antistrange
asymmetry is able to explain a significant part of the NuTeV
anomaly by using both the light-cone baryon-meson fluctuation
model\cite{bm97} and the chiral model
model\cite{Weinberg,Manohar-Georgi}, based on the collaborated
works with Ding\cite{dm04} and also with Ding and
Xu\cite{dxm04,dxm05}.

In the light-cone formalism,
%\cite{ck},
the hadronic wave function can be expressed by a series of
light-cone wave functions multiplied by the Fock states, for
example, the proton wave function can be written as
\begin{eqnarray}
   \left|p\right\rangle=&\left|uud\right\rangle\Psi_{uud/p}+ \left|uudg\right\rangle\Psi_{uudg/p}
\nonumber
\\
   &+\sum_{q\bar{q}}
   \left|uudq \bar{q}\right\rangle\Psi_{uudq\bar{q}/p}+\cdots.
\end{eqnarray}
Brodsky and I made an approximation\cite{bm97}, which suggests
that the intrinsic sea part of the proton function can be
expressed as a sum of meson-baryon Fock states. For example:
$P(uuds\bar{s})=K^{+}(u\bar{s})+\Lambda(uds)$ for the intrinsic
strange sea, the higher Fock states are less important, the $ud$
in $\Lambda$ serves as a spectator in the quark-spectator
model\cite{ma}. The momentum distribution of the intrinsic $s$ and
$\bar{s}$ in the $K^{+}\Lambda$ state can be modelled from the
two-level convolution formula:
\begin{eqnarray}
   s(x)&=&\int^{1}_{x}\frac{\d y}{y}f_{\Lambda/K^{+}\Lambda}(y)q_{s/\Lambda}(x/y),\nonumber\\
   \bar{s}(x)&=&\int^{1}_{x}\frac{\d y}{y}f_{K^{+}/K^{+}\Lambda}(y)q_{\bar{s}/K^{+}}(x/y),
\end{eqnarray}
where $f_{\Lambda/K^{+}\Lambda}(y)$, $f_{K^{+}/K^{+}\Lambda}(y)$
are the probabilities of finding $\Lambda, K^{+}$ in the
$K^{+}\Lambda$ state with the light-cone momentum fraction $y$,
%for the Gaussian type:
%\begin{eqnarray}
%  f_{\Lambda/K^{+}\Lambda}(y)&=&\int^{+\infty}_{-\infty}\d\mathbf{k}_{\bot}\bigg{|}A_{D}
%  \exp[-\frac{1}{8\alpha^{2}_{D}}(\frac{m^{2}_{\Lambda}+\mathbf{k}^{2}_{\bot}}{y}+\frac{m^{2}_{K^{+}}+\mathbf{k}^{2}_{\bot}}{1-y})]\bigg{|}^{2},\nonumber\\
% f_{K^{+}/K^{+}\Lambda}(y)&=&\int^{+\infty}_{-\infty}\d\mathbf{k}_{\bot}\bigg{|}A_{D}
%  \exp[-\frac{1}{8\alpha^{2}_{D}}(\frac{m^{2}_{K^{+}}+\mathbf{k}^{2}_{\bot}}{y}+\frac{m^{2}_{\Lambda}+\mathbf{k}^{2}_{\bot}}{1-y})]\bigg{|}^{2},
%\end{eqnarray}
and $q_{s/\Lambda}(x/y)$, $q_{\bar{s}/K^{+}}(x/y)$ are the
probabilities of finding $s$, $\bar{s}$ quarks in $\Lambda, K^{+}$
state with the light-cone momentum fraction $x/y$.
% for the
%Gaussian type:
%\begin{eqnarray}
%  q_{s/\Lambda}(x/y)&=&\int^{+\infty}_{-\infty}\d\mathbf{k}_{\bot}\bigg{|}A_{D}
%  \exp[-\frac{1}{8\alpha^{2}_{D}}(\frac{m^{2}_{s}+\mathbf{k}^{2}_{\bot}}{x/y}+\frac{m^{2}_{D}+\mathbf{k}^{2}_{\bot}}{1-x/y})]\bigg{|}^{2},\nonumber\\
%  q_{\bar{s}/K^{+}}(x/y)&=&\int^{+\infty}_{-\infty}\d\mathbf{k}_{\bot}\bigg{|}A_{D}
%  \exp[-\frac{1}{8\alpha^{2}_{D}}(\frac{m^{2}_{\bar{s}}+\mathbf{k}^{2}_{\bot}}{x/y}+\frac{m^{2}_{q}+\mathbf{k}^{2}_{\bot}}{1-x/y})]\bigg{|}^{2}.
%\end{eqnarray}
Two wave function models, the Gaussian type and the power-law
type, are adopted\cite{bm97} to evaluate the asymmetry of
strange-antistrange sea, and almost identical distributions of
$s$-$\bar{s}$ are obtained in the nucleon sea. Thus, using this
model, we can obtain the distributions of $s$ and $\bar{s}$ in the
nucleon sea\cite{dm04}. The result of our calculation is
0.0042$<S^{-}<$0.0106 (0.0035$<S^{-}<$0.0087) for the Gaussian
wave function (for the power-law wave function), which corresponds
to $P_{K^{+}\Lambda}$=4\%, 10\%. Hence, $0.0017<\delta
R^{-}_{s}<0.0041$ ($0.0014<\delta R^{-}_{s}<0.0034$), for the
Gaussian wave function (the power-law wave function). The shift in
$\sin^{2}\theta_{w}$ can reduce the NuTeV discrepancy from 0.005
to 0.0033 (0.0036) ($P_{K^{+}\Lambda}$=4\%) or 0.0009 (0.0016)
($P_{K^{+}\Lambda}$=10\%). Thus the $s$-$\bar{s}$ asymmetry can
remove the NuTeV anomaly by about 30--80\% in this model.

%\begin{figure}%[htb]
%\begin{center}
%\includegraphics[width=9.5cm]{Graph2.EPS}
%\caption{\small Distributions for $x\delta_{s}(x)$, with
%$\delta_{s}(x)$=$s(x)-\bar{s}(x)$. The solid curve is for the
%power-law wave function and the dashed curve is for the Gaussian
%wave function.}\label{dssbar}
%\end{center}
%\end{figure}

 A~~further study by Ding, Xu and I\cite{dxm04} by using chiral quark model also shows that this
strange-antistrange asymmetry has a significant contribution to
the PW relation and can explain the anomaly without sensitivity to
input parameters. The chiral symmetry at high energy scale and it
breaking at low energy scale are the basic properties of QCD. The
chiral quark model, established by Weinberg\cite{Weinberg}, and
developed by Manohar and Georgi\cite{Manohar-Georgi}, has been
widely accepted by the hadron physics society as an effective
theory of QCD at low energy scale. This model has also a number of
phenomenological applications, such as to explain the light-flavor
sea asymmetry of $u$ and $d$ sea quarks\cite{ehq92}, and also to
understand the proton spin problem\cite{cl95}. In the new
analysis, we provide a new success to understand the NuTeV anomaly
with the chiral quark model without sensitivity on parameters. We
find that the effect due to strange-antistrange asymmetry can
bring a significant contribution to the NuTeV anomaly of about
60--100\% with reasonable parameters without sensitivity to
different inputs of constituent quark distributions. This may
imply that the NuTeV anomaly can be considered as a
phenomenological support to the strange-antistrange asymmetry of
the nucleon sea. There also similar studies\cite{strange} which
give similar conclusion as ours.

\begin{figure}[htbp]
\begin{center}
%\scalebox{0.78}{\includegraphics[0,16][310,250]{ssbar.EPS}}
\scalebox{0.75}{\includegraphics[0,16][310,250]{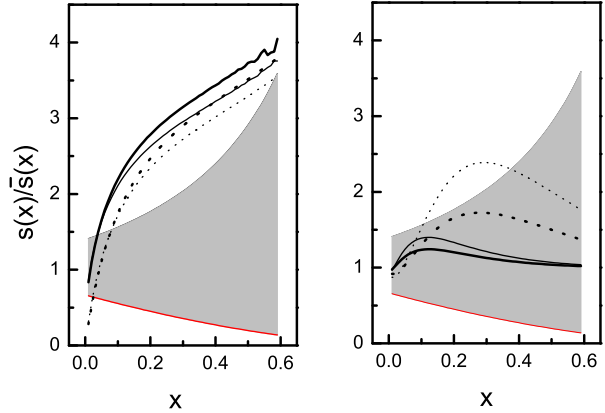}}
\end{center}
\caption[*]{\small Distributions of $s(x)/\bar{s}(x)$, where the
shadowing area is the error range of NuTeV. The thick and thin
curves are the effective chiral quark model results with diferent
inputs. The left side is the prediction by the effective chiral
quark model only and the right side is the result by including
both the prediction of the effective chiral quark model and the
symmetric sea contribution estimated by the difference between the
NuTeV data parametrization and the effective chiral quark model
result.}\label{ssbar}
\end{figure}

Besides, we also calculated\cite{dxm05} the strange sea
distributions of $x(s(x)+\bar{s}(x))$ and $x(s(x)-\bar{s}(x))$
within this model, and notice that the results of the effective
chiral quark model calculations are lower than the parametrization
of NuTeV data\cite{NUTEV} at arbitrary $x$, which may be caused by
the non-considered symmetric strange sea content. This means that
there should be a significant symmetric $s$/$\bar{s}$ contribution
which is not included in the model calculation. We find that the
distribution of $s(x)/\bar{s}(x)$ matches well with the
experimental data\cite{NUTEV}, when additional symmetric sea
contributions being considered effectively by taking into account
the difference between model results and data parametrization.
Thus the calculated $s(x)/\bar{s}(x)$  asymmetry are compatible
with the data by including some additional symmetric strange quark
contribution, as can be seen from Fig.~\ref{ssbar}.

Gao and I also analyzed\cite{gm05} the possible light-quark
fragmentation effect from prompt like-sign dimuon data and studied
its influence on the measurement of strange asymmetry by
NuTeV\cite{NUTEV}. Our result is that the light-quark
fragmentation may be an important source that reduces the effect
of strange asymmetry from opposite sign dimuon
studies\cite{NUTEV}. The difference for the $D(c\bar{q})$ and
$\bar{D}(\bar{c}q)$ meson production cross sections in neutrino
and antineutrino induced charged current deep inelastic scattering
is illustrated to be sensitive to the nucleon strange
asymmetry\cite{gm05}. There is also a suggestion to measure the
strange asymmetry by $D_s$ asymmetry in
photoproduction\cite{qiao}.

Finally, we give our conclusions as follows:
\begin{itemize}
\item
The
effect due to strange-antistrange asymmetry might be important to
explain the NuTeV anamoly or the NuTeV anomaly could be served as
an evidence for the $s/\bar{s}$ asymmetry.
\item
The calculated $s/\bar{s}$ asymmetry are compatible with the
available data by including some additional symmetric strange
quark contribution.
\item
Reliable precision measurements are needed to make a crucial test
of $s/\bar{s}$ asymmetry. \end{itemize}

{\bf Acknowledgments} This work is partially supported by National
Natural Science Foundation of China (Nos.~10421003 and 10575003),
and by the Key Grant Project of Chinese Ministry of Education
(No.~305001).

\end{document}